\documentclass[12pt,preprint]{aastex}
\usepackage{amsmath}
\usepackage{graphicx}% Include figure files
\usepackage{dcolumn}% Align table columns on decimal point
\usepackage{bm}% bold math

\shorttitle{Planet Formation around Intermediate Mass Stars}
\shortauthors{Kretke et al.}

\newcommand{\Mdot}{\dot{M}}
\newcommand{\md}{{\rm mid}}
\newcommand{\St}{{\rm St}}
\newcommand{\Stmax}{{\rm St_{max}}}
\newcommand{\OmK}{\Omega_{\rm K}}
\newcommand{\aeff}{\alpha_{\rm eff}}
\newcommand{\aMRI}{\alpha_{\rm MRI}}
\newcommand{\acrit}{a_{\rm crit}}
\newcommand{\yr}{{\rm yr}}

\newcommand{\smax}{s_{\rm max}}
\newcommand{\sint}{\displaystyle\int_{s_{\rm min}}^{s_{\rm max}}}
\newcommand{\Miso}{M_{\rm iso}}
\newcommand{\AU}{{\rm AU}}

\begin{document}
\title{Assembling the Building Blocks of Giant Planets around 
Intermediate Mass Stars}
\author{K. A. Kretke$^1$, D. N. C. Lin$^{1,2}$,  P. Garaud$^{3}$, N. J. Turner$^{4}$}
\affil{$^1$Department of Astronomy and Astrophysics, University
of California, Santa Cruz, CA, 95064}
\affil{$^2$Kavli Institute of Astronomy and Astrophysics, Peking
University, Beijing, China}
\affil{$^3$Department of Applied Mathematics and Statistics, Baskin School of Engineering, University of California, Santa Cruz, CA, 95064}
\affil{$^4$ Jet Propulsion Laboratory, California Institute of Technology, Pasadena, CA, 91109}
\email{kretke@ucolick.org}

\begin{abstract}
We examine a physical process that leads to the efficient formation of gas giant planets around intermediate mass stars. In the gaseous protoplanetary disks surrounding rapidly-accreting intermediate-mass stars we show that the midplane temperature (heated primarily by turbulent dissipation) can reach $\gtrsim$1000~K out to 1 AU. Thermal ionization of this hot gas couples the disk to the magnetic field, allowing the magneto-rotational instability (MRI) to generate turbulence and transport angular momentum.  Further from the central star the ionization fraction decreases, decoupling the disk from the magnetic field and reducing the efficiency of angular momentum transport.  As the disk evolves towards a quasi-steady state, a local maximum in the surface density and in the midplane pressure both develop at the inner edge of the MRI-dead zone, trapping inwardly migrating solid bodies. Small particles accumulate and coagulate into planetesimals which grow rapidly until they reach isolation mass.  In contrast to the situation around solar type stars, we show that the isolation mass for cores at this critical radius around the more massive stars is large enough to promote the accretion of significant amounts of gas prior to disk depletion.  Through this process, we anticipate a prolific production of gas giants at $\sim 1$ AU around intermediate-mass stars.
\end{abstract}
\keywords{planetary systems: formation, protoplanetary disks}

\section{Introduction}
The discovery of a plethora of extra-solar planets around solar-type main
sequence stars has established that planet formation must be a common process, not a peculiarity of our own solar system.
As observational techniques for planetary detection have become
more sophisticated, the discovery domain has expanded to include host
stars with a wide range of masses.  
While on the main sequence, intermediate mass stars (stars with $1.5
M_{\sun} \lesssim M_* \lesssim 3 M_{\sun}$) make poor radial velocity
(RV) survey candidates as they have few spectral lines which also 
tend to be rotationally broadened (\citet{Griffin.etal.2000} but see \citet{Galland.etal.2006}). However, once these
stars evolve off the main sequence, their relatively cool and slowly
rotating outer layers make them more suitable candidates for high
precision spectroscopic studies.
Recent RV surveys targeting evolved intermediate mass stars suggest that  
they differ from solar-type stars as planetary hosts in at least two respects.
First, the total frequency of giant planets (with periods less than a few years) appears be higher around
intermediate mass stars. Second, the planets have different statistical properties.  Their semi-major axis distribution is
concentrated at 1-2 AU and there is an apparent
lack of short-period (days to months) planets, despite observational
selection effects favoring their discovery \citep{LovisMayor.2007}.

In this paper, we propose a common explanation for prolific
gas giant formation with semi-major axes comparable to 1~AU
and for the rarity of close-in planets around intermediate-mass
stars.  As it is unlikely that all planets within 1 AU have been engulfed or had their orbits disrupted
by the current expanded envelope of the host stars \citep{Johnson.etal.2007}, 
we attribute both properties to the formation and early evolutionary
processes rather than to post-main-sequence evolution.

We begin by examining the physical properties of
circumstellar disks which may affect the probability of forming giant
planets. In the core-accretion model of planet formation
(cf. \citet{Bodenheimer.Pollack.1986}), the emergence of Jupiter-like
gas giants requires that a population of solid cores form
within a gaseous protoplanetary disk. These cores grow
through cohesive collisions with planetesimals,
with a growth rate determined by the velocity dispersion
of the planetesimal swarm.  
The magnitude of this velocity dispersion is set by a balance between excitation by gravitational
perturbations and damping by gas drag. In the
gas-rich environment of typical protostellar disk, 
gas drag dominates so that field planetesimals only attain
relatively small equilibrium velocity dispersion. As a result the most
massive protoplanetary embryos can access only those building blocks within
their gravitational feeding zones \citep{Kokubo.Ida.1998}.  When these embryos have collected all the
planetesimals within about five times their Roche radius on
either side of their orbits, their growth stalls. 
This maximum embryo mass,
a function of planetesimal surface density and distance from the central star,
is referred to as the embryo's isolation mass ($\Miso$).  
Gas giants can only form if the
embryos' $\Miso$ is sufficiently large for the cores to  
begin accreting gas prior to the depletion of their nascent disks \citep{IdaLin2004}.

Although the gravity of lunar-mass embryos is adequate to accrete disk gas
with temperature $<10^3$ K, \emph{efficient} dynamical gas accretion 
is only possible for cores with masses greater than
some critical value ($M_{\rm crit}$). In a minimum mass solar nebula \citep{Hayashi.1981} with an interstellar
grain size distribution, $M_{\rm crit} \sim 10 M_\oplus$ at a semi-major axis $a\sim
5$~AU \citep{Pollack.etal.1996}, although this critical mass decreases both
with lowered grain opacity \citep{Ikoma.etal.2000, Hubickyj.etal.2005} and with increased density of the
ambient gas \citep{Bodenheimer.Pollack.1986, Papaloizou.Terquem.1999}.  
Gas giant formation therefore requires that the 
heavy-elements in the disk can be efficiently assembled into massive cores with mass greater than $M_{\rm crit}$.
In order to understand the spatial
distribution of the gas giant planets we must understand
how the building blocks of these cores migrate and are retained in gaseous disks.

In protoplanetary disks solid retention first  becomes an issue once grains grow beyond a few cm in size.
In most regions of protostellar disks, the midplane pressure ($P_{\rm
mid}$) decreases with distance from the central star ($r$) so that the
gas is slightly pressure supported, resulting in a sub-Keplerian
azimuthal velocity. Grains larger than a few cm are decoupled from the
gas and move at Keplerian speeds.  Consequently, grains typically experience head
winds and undergo orbital decay \citep{Weidenschilling1977}.  However
if $P_{\rm mid}$ does not monotonically decrease with $r$ then immediately
interior to a local pressure maximum the gas attains
super-Keplerian velocities. This motion introduces a tail wind on
the decoupled grains and causes them to drift outwards towards local pressure
maxima \citep{Bryden.etal.2000, HaghihipourBoss2003}.

Solid retention again becomes an issue once 
planetesimals grow into earth-mass embryos and tidal interactions
with the gaseous disk become important.  Before embryos are sufficiently massive to
open up gaps in the disks \citep{LinPapaloizou1986}, they can exchange angular momentum
with the gas via their Lindblad and
co-rotation resonances \citep{Goldreich.Tremaine.1980}.  A geometric bias causes an
imbalance between the Lindblad resonances which generally leads to a loss of
angular momentum and orbital decay for the embryos \citep{Ward.1986,
Ward1997}. 
However, embryos will gain angular momentum through their
co-rotation resonances if there is a positive $P_{\rm mid}$ gradient
\citep{Tanaka.etal.2002, Masset.etal.2006}. 
Numerical models which take into account
these physical effects have reproduced the observed $M_p-a$
distribution around solar type stars \citep{IdaLin2008}.

Several physical processes can lead to local maxima in $P_{\rm mid}$.
Various authors have explored the potential accumulation of grains at
transient pressure maxima formed by turbulent fluctuations \citep{Johansen.etal.2006b,Fromang.Nelson.2005} or spiral waves \citep{Rice.etal.2006}.
These mechanisms, while likely extremely important for forming planetesimals at a large range of radii, are still quite ``leaky'' as a significant fraction of the solid material simply undergoes a slightly slower random walk towards the central star.
However, longer lived pressure maxima may also exist due to large scale changes in the disk viscosity \citep{Kretke.Lin.2007}.

We expect radial variations in viscosity if turbulence caused by the magneto-rotational instability (MRI; \citet{BalbusHawley1991}) is the primary mechanism for transporting angular momentum.
These variations result from changes in the ionization fraction at different radii in the disk since free electrons are needed to couple the gas to the magnetic field.
The disk is thermally ionized in the hot inner regions, but further out
stellar x-rays and diffuse cosmic rays ionize only 
the surface layers, resulting in a viscously active
turbulent surface sandwiching an inactive ``dead zone''
\citep{Gammie1996}. 
At the critical radius marking the inner edge of the dead zone ($\acrit$), the effective viscosity decreases with increasing distance from the central star.  
In a quasi-steady state situation (expected to develop rapidly in the inner regions of the disk) this decrease in viscosity leads to a local increase
in the magnitude of $\Sigma_g$ and hence of $P_{\rm mid}$ 
with radius.  This disk structure provides a promising
barrier to the orbital decay of both boulders and embryos.
The radial location of $\acrit$ depends on the stellar mass and the mass accretion rate which we argue explains the observed differences between the statistical distributions of planets around solar-type stars and around intermediate mass stars.

In this paper, we present a model for the formation of
planets at the inner edge of the dead zone and argue why this process is more relevant for
intermediate mass than for solar mass stars.  
In $\S$\ref{sec:model} we describe our quantitative numerical model for the evolution of solids in the disk, based upon the work of \citet{Garaud.2007}.
An important aspect of this model is the location of $\acrit$ (the inner edge of the dead zone)  which we derive in $\S$\ref{sec:acrit} as a function of stellar mass and mass accretion rate.
In $\S$\ref{sec:results} we present the model results for a $2~M_\sun$ star and estimate how this planet formation mechanism scales with stellar mass.
In $\S$\ref{sec:summary} we summarize our conclusions.

\section{Model Description}\label{sec:model}
In order to assess the probability of forming of gas giants at the
inner edge of the dead zone, we must calculate the expected isolation
mass of the cores ($\Miso$) at this location ($\acrit$).  
From the work of \citet{Kokubo.Ida.1998}, it has been established that the embryo's isolation mass is sensitive both to the distance from the host star and to the surface density of solids. 
For the purpose of computing the efficiency of solid retention we calculate the 
 dynamical evolution of the solids in the disk using a 
modification of the numerical scheme developed by \citet{Garaud.2007} (hereafter G07).
For full details we refer the readers to G07 but
for reference the salient points are described here.  In this model we
assume that particles, as a result of a collisional cascade, maintain
a power-law size distribution in which the number density of particles of size $s$ goes as $dn/ds
\propto s^{-3.5}$, with sizes ranging from a fixed $s_{\rm min}$ to
$s_{\rm max}$ and where $s_{\rm max}$ is allowed to vary with time and distance from the central star.  This simplifying
assumption allows us to completely describe the evolution of the gas
and solids in the disk by the total gas
surface density ($\Sigma_g(r,t)$), the solid and vapor surface densities
of the different species in the disk ($\Sigma_{p,i}(r,t)$ and
$\Sigma_{v,i}(r,t)$ respectively), and the maximum size of the particles
$s_{\rm max}(r,t)$.

In the following sections we describe how we have modified the G07 scheme.
In \S\ref{sec:vis}, we describe how we
account for the presence of an evolving dead zone.
In particular we describe in \S\ref{sec:acrit} how we calculate the location of the inner edge of the dead zone ($\acrit$) using a simple model of the vertical thermal structure.
Finally, in \S\ref{sec:feedback} we model the drag on the gas caused by the
particles in the limit of large $Z=\Sigma_p/\Sigma_g$, an effect important near $\acrit$ where solids are seen to accumulate.

\subsection{Viscosity at a function of $r$ and $t$}
\label{sec:vis}
The original G07 model uses the standard $\alpha$ viscosity prescription
($\nu(r)=\aeff c_s(r) h(r)$) where $c_s$ is the midplane sound-speed, $h$ is the disk scale-height ($h\equiv c_{s,{\rm mid}} \OmK^{-1}$) and $\aeff$ is constant.  In this paper we consider $\aeff = \aeff(r,t)$ to model the presence of an evolving MRI-dead zone.
In fully MRI-active regions (eg. interior to $\acrit$) we assume that $\aeff=\alpha_{\rm MRI}$, while exterior to $\acrit$, $\aeff$ is modified to take into account the lower viscosity of the dead zone.
In \S\ref{sec:acrit} we describe in detail how we calculate the location of $\acrit$.

Exterior to $\acrit$ the disk is ionized by x-rays and cosmic rays, so only the surface layers are MRI active.  The column density of this active layer ($\Sigma_A$) is strongly dependent on the ionizing source (ie. the ability of cosmic rays to penetrate the stellar magnetosphere) and on the recombination rate, which is dominated by the amount of small grains.  Due to these uncertainties, and the fact that the exact evolution of the outer parts of the dead zone are relatively unimportant in these calculations, we follow previous studies and assume that $\Sigma_A = 100$~g cm$^{-2}$  at all radii
(outside of the thermally ionized region) (eg. \citet{Gammie1996}).

Additionally, we assume that there is some amount of angular momentum transport in the dead zone.  
This transport may be due to the propagation of MRI-driven waves into the laminar dead zone
(\cite{Fleming.Stone.2003,Turner.etal.2007}) or to a mechanism unrelated to the MRI. 
This motivates the following prescription in regions beyond $\acrit$
\begin{equation}
\aeff(r,t) = 
 \begin{cases} 
       \frac{2\alpha_{\rm MRI}\Sigma_A + \alpha_{ \rm eff, dead} (\Sigma_g - 2\Sigma_A)}{\Sigma_g (r,t)},&~{\rm if}~ \Sigma_g \ge 2 \Sigma_A,\\
       \alpha_{\rm MRI},&~{\rm otherwise},
  \end{cases}     
\label{eq:aeff}
\end{equation}
where $\alpha_{ \rm eff, dead}$ is the effective viscosity in the dead zone.  

\subsection{Location of $\acrit$}
\label{sec:acrit}
As the degree of thermal ionization is a sensitive function of temperature, we must calculate the vertical structure of the disk in order to find the location of $\acrit$, the inner edge of the dead zone.
In the innermost regions of the disk viscous heating dominates the energy budget, and stellar irradiation can be neglected \citep{GaraudLin2007}. 

We assume that the disk is in hydrostatic equilibrium
\begin{equation}
\frac{d P}{dz} = -\rho_g \Omega_K^2 z,
\end{equation}
and is viscously heated so that the vertical energy flux ($F$) is described by
\begin{equation}
\frac{d F}{dz} = \frac{9}{4}\nu\rho_g\Omega_K^2.
\end{equation}
Here we consider that $\nu(r,z)$  is a function of both height and radius in the disk and assume that the orbital frequency can be approximated by the Keplerian frequency.

We also assume radiative energy transport so that
\begin{equation}
F = -\frac{4 a c}{3}\frac{T^3}{\kappa \rho_g} \frac{dT}{dz},
\end{equation}
where $\kappa$ is the Rosseland mean opacity.
For the temperature range of interest (400-1,200K) the dominant
opacity sources are silicate and iron grains.  We adopt a grey
opacity approximation in which $\kappa=1~\rm{cm^2~g^{-1}}$.  This
approximation is consistent with the \citet{Ferguson.etal.2005} dust
opacities.  
We solve only the optically thick region of the disk and use photospheric boundary conditions at $z=z_e$, namely $P(z_e) = (2/3)\Omega_K^2 z_e/\kappa$ and $F(z_e)=\sigma T(z_e)^4$. Finally we assume symmetry about the midplane.

In order to parametrize the variation of the viscosity with height above the midplane we follow the results of 3D MHD simulations which demonstrate that, in fully developed MRI-turbulence, the shear stress($w\equiv \nu\rho_g r (d\Omega/dr) = (3/2) \nu \rho_g \OmK$) is approximately constant with height \citep{MillerStone2000}.   Based on these numerical
results, we model the viscosity at a given radius as
\begin{equation}
\nu(z) = 
    \begin{cases} \frac{2}{3}\frac{\alpha P_\md}{\rho_g(z) \OmK}, & {\rm if}~|z| < z_\nu, \\
                  0 ,& \rm{otherwise},
     \end{cases} \label{eq:nu}
\end{equation}
where $z_\nu \approx 2h$.
Note that this differs from the more common 2D parametrization sometime referred to as the ``$\alpha P$-formalism'' (e.g. \citet{Cannizzo1992}) where $\nu_{\alpha-P}=\alpha c_s^2(z) \OmK^{-1}$.  Viscosity described by equation (\ref{eq:nu}) increases with height, which leads to a cooler midplane more consistent with multi-dimensional MHD simulations than the $\alpha P$-formalism \citep{Hirose.etal.2006}.

Equation (\ref{eq:nu}) is applicable to MRI-active regions 
where the partially ionized gas is well coupled to the
turbulent magnetic field through the entire thickness of the disk.
As long as the midplane temperature is high enough to sufficiently ionize the gas this structure is self-consistent. We can therefore use it to determine $\acrit$, the outermost radius where this condition is satisfied.  
In practice we combine the equations for the vertical structure 
with the Saha equation to calculate the location at which the midplane just satisfies the
ionization criteria of $x_e \ge 10^{-12}$ (where $x_e$ is the fractional number density of electrons) which corresponds to $T
\sim 1000$K for typical disk midplane densities \citep{Umebayashi1983}.

In Figure \ref{acrit} the curves show the location of the inner edge
of the dead zone as a function of stellar mass and accretion rate
assuming that in the active region $\alpha=10^{-2}$.  
These theoretical curves can be approximated by
\begin{eqnarray}
\acrit &=& 0.77 \left(\frac{\Mdot}{1.4 \times 10^{-7} M_\sun
\rm{yr}^{-1}}\right)^{4/9} \left(\frac{M_*}{2 M_\sun}\right)^{1/3}\nonumber\\
&&\times\left(\frac{\alpha}{10^{-2}}\right)^{-1/5}
\left(\frac{\kappa_D}{1~{\rm cm^2~g^{-1}}}\right)^{1/4}{\rm AU},
\label{eq:acrit}
\end{eqnarray}
where the quasi-steady-state approximation for the mass accretion rate
\begin{equation}
\dot{M} = 3\pi\displaystyle\int_{-\infty}^{\infty}\rho_g\nu dz =  4\pi\alpha P_{\rm mid} \OmK^{-1} z_\nu
\end{equation}
was used to eliminate $P_{\rm mid}$.  
We find that this value for $\acrit$ varies significantly from the $\sim 0.1$~AU often quoted (ie. \citet{Gammie1996}) if the stellar mass or the mass accretion rate are large.
By using the relationship 
\begin{equation}
\Mdot = 3\pi\nu\Sigma_g = 3\pi \aeff c_s^2 \OmK^{-1} \Sigma_g \label{eq:Mdot_general}
\end{equation}
\citep{Pringle1981} we can relate the $\alpha$ in equation (\ref{eq:acrit}) to the the vertically averaged $\aeff$ when the disk is fully MRI active and find that $\alpha_{\rm MRI} \approx \alpha$.

Observationally, $\Mdot$ appears to be correlated with $M_*$, although the exact relationship is both uncertain and shows significant
scatter.  For reference the symbols in Figure 2 indicate measurements
of mass accretion rates onto young stars.  The solid points are from
observations of the young cluster ($<$ 1 Myr) $\rho$-Oph by \citet{Natta.etal.2006} and the dashed line
shows the best fit to their data which is
\begin{equation}
\dot{M}\simeq 4 \times 10^{-8} (M_*/ M_\sun)^{1.8} M_\sun 
{\rm yr}^{-1}.
\label{eq:mdot}
\end{equation}
As the mass accretion rate for the higher mass stars in the Natta et al. sample is dominated by a
single object, we have also plotted (as open points) similar data from older,
heterogeneously distributed intermediate mass stars
with estimated ages ranging from 1-10~Myr \citep{GarciaLopez.etal.2006}. As suggested in this study, these systematically older stars may have had higher
mass accretion rates consistent with the extrapolation from
$\rho$-Oph, when they were younger.

Using the best fit relationship from $\rho$-Oph implies that $a_{\rm
crit} \propto M_*^{1.1}$.  \citet{ClarkePringle2006} suggest that the
correlation between stellar mass and mass accretion rate may not be
this steep due to observational biases.  If instead we use their
estimation that $\Mdot \propto M_*$, then $a_{\rm crit}\propto M_*^{0.8}$,
which is a slightly less sensitive but nevertheless increasing function
of $M_*$.  Therefore, in the light of these the uncertainties, we will simply assume that $\acrit \propto M_*$. 

\subsection{Gas-Particle Feedback}
\label{sec:feedback}
In a thin disk, the global evolution of the gas surface density
$\Sigma_g$ is determined by the equation
\begin{equation}
\frac{\partial \Sigma_g}{\partial t}
+ \frac{1}{r} \frac{\partial}{\partial r} \left(\Sigma_g u_r r\right) = 0.
\label{eq:diffuse}
\end{equation}
In a standard viscous accretion disk the radial velocity ($u_r$) of gas equals $u_\nu$ where 
\begin{equation}
u_\nu\equiv-\frac{3}{r^{1/2}\Sigma_g}\frac{\partial}{\partial r}(r^{1/2}
\nu \Sigma_g). \label{eq:v_nu}
\end{equation}
In our analysis, we consider the possibility that the radial velocity
of the gas $u_r$ is not only determined by the viscosity but also by
momentum transferred via drag between the gas and the solid particles.  
We calculate the equation of motion for a
parcel of gas and dust in a similar fashion to \citet{Nakagawa.etal.1986}. 
However, instead of assuming a single particle size for the solids we consider the power-law distribution ($dn/ds \propto s^{-3.5}$).
The equation of motion for a given particle of size $s$ is
\begin{equation}
\frac{d {\bm V}(s)}{d t} = -\frac{1}{\tau_s(s)}({\bm V}(s)- {\bm U}) -\frac{G
  M_*}{r^3}{\bm r}\label{eq:vel1}
\end{equation}
where $\tau_s(s)= s \rho_s/(\rho_g c_s)$ is the stopping time for a particle of size $s$ and density $\rho_s$ (in the Epstein regime, see G07 for a full description), where ${\bm V}(s)$ is the size dependent particle velocity and $\bm U$ is the gas velocity.

Assuming that deviations from Keplerian orbital velocity are small, we write ${\bm V}(s)=v_r(s){\bm\hat{r}} + (r\OmK+v_{\phi}(s)){\bm{\hat{\phi}}}$ and ${\bm U}=u_r{\bm\hat{r}} + (r\OmK+u_{\phi}){\bm{\hat{\phi}}}$). The linearization of equation (\ref{eq:vel1}) yields
\begin{eqnarray}
\frac{\partial v_r}{\partial t} &=& -\frac{1}{\tau_s(s)}(v_r(s)-u_r)+2\OmK
v_\phi(s),\label{eq:vr}\nonumber\\ 
\frac{\partial v_\phi}{\partial t} &=&
-\frac{1}{\tau_s(s)}(v_\phi(s)-u_\phi)-\frac{1}{2}\OmK v_r(s),\label{eq:vphi}
\end{eqnarray}
The particles and gas adjust to a steady motion with respect
to each other within a few stopping times, so we solve for the steady-state solutions only. 
Multiplying by the particle mass and integrating over the whole size distribution function yields
\begin{eqnarray}
\sint\left[-m(s)\frac{dn}{ds}\frac{1}{\tau_s(s)}(v_r(s)-u_r)+2\OmK m(s)\frac{dn}{ds}v_\phi(s)\right]ds &=& 0\label{eq:mm_p1},\nonumber\\
\sint\left[-m(s)\frac{dn}{ds} \frac{1}{\tau_s(s)}(v_\phi(s)-u_\phi)-\frac{1}{2}\OmK m(s)\frac{dn}{ds}v_r(s)\right]ds &=& 0\label{eq:mm_p2}.
\end{eqnarray}
The momentum lost by the solids is acquired by the gas, so that the steady state gas-dynamic can be described by
\begin{eqnarray}
\sint m(s)\frac{dn}{ds}\frac{1}{\tau_s(s)}(v_r(s)-u_r)ds +2\OmK\rho_g(u_\phi+\eta v_K) &=& 0\label{eq:mm_g1},\nonumber\\
\sint m(s)\frac{dn}{ds}\frac{1}{\tau_s(s)}(v_\phi(s)-u_\phi)ds-\frac{1}{2}\OmK\rho_g(u_r-u_\nu) &=& 0\label{eq:mm_g2},
\end{eqnarray}
where $\eta$ is the non-dimensional pressure gradient
\begin{equation}
\eta \equiv -\frac{1}{2}\frac{h^2}{r^2}\frac{\partial \ln P}{\partial \ln r},
\end{equation}
and $v_K=r \OmK$.
In most regions of the disk $\eta$
is positive due to the negative pressure gradient, but its sign will change around pressure maxima.
In the absence of particles, the gas will have an
azimuthal velocity (relative to the Keplerian motion) of $u_\phi =
-\eta v_K$ and a radial velocity of $u_r = u_\nu$.

Combining these equations and
defining the mass--weighted average particle velocities as
\begin{eqnarray}
\bar{v}_\phi &\equiv& \frac{1}{\rho_p}\sint m(s) \frac{dn}{ds}v_\phi(s)ds,\nonumber\\
\bar{v}_r &\equiv& \frac{1}{\rho_p}\sint m(s) \frac{dn}{ds}v_r(s)ds
\end{eqnarray}
allows us to write
\begin{eqnarray}
u_\phi &=& -\frac{\rho_p}{\rho_g} \bar{v}_\phi \label{eq:1},\nonumber\\
u_r &=& u_\nu - \frac{\rho_p}{\rho_g} \bar{v}_r \label{eq:2}.
\end{eqnarray}
Combining this with equation (\ref{eq:vphi}) we can solve for
\begin{eqnarray}
v_r(s) &=& \frac{u_r+2\OmK\tau_s(s) u_\phi}{1+\OmK^2\tau_s(s)^2},\nonumber\\
v_\phi(s) &=&\frac{u_\phi-(1/2)\OmK\tau_s(s)u_r}{1+\OmK^2\tau_s(s)^2}.
\end{eqnarray}
Using the definition for the Stokes number $\St(s)=\OmK\tau_s(s)/(2\pi)$ and taking the mass--weighted integral over all particle sizes yields
\begin{eqnarray}
\bar{v}_r = I\left(\sqrt{2\pi \Stmax}\right)u_r+2 J\left(\sqrt{2\pi \Stmax}\right)u_\phi,\label{eq:3}\nonumber\\
\bar{v}_\phi = I\left(\sqrt{2\pi \Stmax}\right)u_\phi-\frac{1}{2} J\left(\sqrt{2\pi \Stmax}\right)u_r\label{eq:4},
\end{eqnarray}
where $I$ and $J$ are the same as equation (52) in G07 (reproduced in Appendix A for reference) and $\Stmax = \St(\smax)$.
Combining equations (\ref{eq:2}) and (\ref{eq:4}) yields the final steady-state velocities for the particles and the gas where $I \equiv I\left(\sqrt{2\pi \Stmax}\right)$, $J \equiv J\left(\sqrt{2\pi \Stmax}\right)$, and $\chi \equiv \rho_p/\rho_g$.
\

\begin{eqnarray}
\bar{v}_r &=& \frac{[I + \chi (I^2+J^2)]u_\nu - 2J\eta v_K}{1+2 \chi I+ \chi^2(I^2+J^2)},\nonumber\\
\bar{v}_\phi &=& -\frac{1}{2}\left[\frac{J u_\nu + 2[I+\chi(I^2+ J^2)]\eta v_K}{1+2 \chi I + \chi^2(I^2+J^2)}\right]\label{eq:part_vel},\\
u_r &=& \frac{(1 + \chi I)u_\nu + 2J\chi\eta v_K}{1+2 \chi I+ \chi^2(I^2+J^2)},~{\rm and}\nonumber\\
u_\phi &=& \frac{1}{2}\left[\frac{\chi J u_\nu - 2(1 + \chi I)\eta v_K}{1+2 \chi I+ \chi^2(I^2+J^2)}\right]\label{eq:gas_vel}.
\end{eqnarray}

In the gas dominated limit ($\chi \rightarrow 0$) these equations reduce to those in G07.
If $\Stmax$ is large and $\Stmax \gg \chi^2$, the solids will be decoupled from the gas and will not migrate significantly
($v_r = 0$) while the gas evolves viscously ($u_r=u_\nu$).  
In the limit $\Stmax \ll 1$, the grains
are well-coupled to the gas so there will be little relative motion between the two ($u_r=v_r=u_\nu/(1+\chi)$).

\section{Model Results}\label{sec:results}
With these modifications to the G07 prescription 
we calculate the evolution of a disk around a $2 M_\sun$ star including an MRI-dead zone
($\alpha_{\rm MRI}=10^{-2}$,~$\alpha_{ \rm eff,dead}=10^{-3}$).
The initial disk has a surface density profile of
\begin{equation}
\Sigma_g = \Sigma_0 \left(\frac{r}{1 \AU}\right)^{-1} \exp\left(\frac{-r}{R_0}\right)
\end{equation}
where $R_0=30 $AU, and $\Sigma_0=10^4~\rm{g~cm^{-2}}$ has been chosen such that the quasi-steady state accretion rate $\Mdot$ from equation (\ref{eq:Mdot_general}) is $1.4\times10^{-7} \Mdot~\rm{yr}^{-1}$ when $\aeff = \alpha_{\rm eff,dead}$.
For the chosen value of $\alpha_{\rm eff,dead}$ the total disk mass is 0.1$M_*$.  This relatively massive disk is still stable according to the Toomre criterion at all radii. 
For simplicity we only track one species of solids, a generic refractory material which we take to be a combination of silicates and metals, materials which are assumed to sublimate at 1500K. 
Using the solar composition from \citet{Lodders2003} we begin with a dust-to-gas ratio of 0.005 for these refractory materials.
We do not track volatile ices as they will not contribute to the solid cores formed in the hot regions of interest.
For other model parameters we use the values presented in the fiducial model of G07.

\subsection{Accumulation of Solids}\label{sec:evolve2}
We first study the accumulation of solids near the edge of the dead zone and emphasize the importance of feedback in terms of momentum exchange between the solids and the gas. 
We present two runs, the first neglecting momentum transfer from the dust to the gas and then including this feedback.

The top and middle panels of figure \ref{fig:sigma} show the evolution of the gas and solids neglecting the feedback of the dust on the gas.
For reference, figure \ref{fig:alpha} shows $\aeff$ at $10^4$ years and at $10^5$ years to indicate the location of the dead zone.  Early on the dead zone extends from 0.9 to 20 AU, but as the disk evolves the dead zone shrinks. 
Within the first $10^4$ years the gaseous disk adjusts to a quasi-steady-state profile in the inner regions.
The surface density of the gas in the fully MRI-active inner region ($r \lesssim$ 1 AU) is reduced by a factor of $\alpha_{\rm eff,dead}/\alpha_{\rm MRI}$ compared with that in the dead zone ($r \in [1,10]$ AU), thus creating a local pressure maximum near 1 AU.
In this calculation neglecting feedback, the solids accumulate in a very narrow ring corresponding to this pressure maximum.
After only $5.3\times10^5$ years the largest-size body at this location reaches $5 M_\earth$ and
the core thus formed is expected to continue to grow significantly due to accretion of gas, not included in this numerical calculation.   Crucially, in this model tidal interactions will not affect the orbital evolution of the core as the effect of Type I migration would be to keep the core at the pressure maximum \citep{Masset.etal.2006}.
It is also interesting to note that when feedback is neglected, virtually all solids are trapped at the pressure maximum and negligible amounts of heavy elements accrete onto the star.
This effect is associated with the clear spatial separation of the pressure maximum and the sublimation line (here at $\sim$ 0.3 AU) and would appear to predict that intermediate-mass stars should be strongly depleted in refractory elements.
However we now show that this is in fact unrealistic as the large build up of material makes it necessary to include the transfer of momentum from the solids to the gas.

The bottom panels of figure \ref{fig:sigma} show the evolution of the same initial disk including this momentum feedback.  The early evolution is essentially similar for $t < 10^4$ yrs. 
However, once the amount of solids in the
inner edge of the dead zone has increased by an order of magnitude, it begins
affecting the gas properties (compare the middle and bottom panels of fig \ref{fig:sigma}).  
The mass accretion rate onto the star decreases as the gas receives angular momentum from the solids. The decrease in $\Mdot$ reduces the surface density and thus the midplane temperature, causing $\acrit$ to move inwards.  
As $\acrit$ moves inwards the particles of different sizes respond to this shift in location at different rates, smoothing the sharp peak in the solid surface density and allowing the trapped gasses to accrete onto the central star.  
In this way both $\Mdot$ and the position of $\acrit$ oscillate with time as seen in figure \ref{fig:Mdot}.
It is also interesting to note that these oscillations allow heavy elements to accrete onto the star since when $\acrit$ moves outwards it leaves some solids interior to the ``trap'' of the pressure maximum.
Thus intermediate-mass stars will not, in fact, be significantly depleted in refractory elements.
In these simulations we find that the total amount of solids trapped near the pressure maximum is of the order of 100 $M_\earth$ and that the dust-to-gas ratio in this region is of order unity.
The core growth is slower than in the simulation without feedback as material is deposited over a wider range of radii.   In $5.3 \times 10^5$ yrs the largest body has grown to $0.5 M_\earth$ but the amount of material in its vicinity suggests that it will continue to grow larger still.

\subsection{Core Formation}
\label{sec:core}
As the core mass increases much beyond 0.5 $M_\earth$ many effects not included in the numerical algorithm could affect its growth.  For example, it may be important to include the effects of tidal interactions since the oscillations of the position of the pressure maximum could affect the orbital evolution of the core.  Additionally, gravitational interactions between solids may start to impact their radial distribution.  The cores may scatter each other out of the region of interest and radially migrating solids will be trapped into resonances with the existing core, stalling radial migration \citep{Weidenschilling.Davis.1985}.  As a result, we do not continue integrating the simulations beyond this point but instead estimate the achievable core mass with simpler but robust scalings based on our most important results from the previous section: (1) the presence of a large reservoir of solids ($>100 M_\earth$) around $\acrit$ and (2) the saturation of the dust-to-gas surface density ratio to $Z\sim1$ near $\acrit$ (as seen in fig. \ref{fig:sigma}), which is a simple consequence of the saturation of the nonlinear momentum feedback between the solids and the gas.  

Using these ideas, we estimate the \emph{maximum} achievable core size from the isolation mass \citep{IdaLin2005} at $\acrit$ is 
\begin{equation}
\Miso = 2\pi \Sigma_p \acrit b r_H(\acrit),
\end{equation}
where $r_H(\acrit) = \acrit (2 \Miso/(3M_*))^{1/3}$ is the Hill's radius and $b\simeq10$ \citep{Lissauer.1987}.  
We assume that $\Sigma_g$ is related to $\Mdot$ by the quasi-steady-state approximation (\ref{eq:Mdot_general}), and that
$\Sigma_p$ is related to the gas density via $Z=\Sigma_p/\Sigma_g$ where $Z\sim1$.  We find that
\begin{eqnarray}
\Miso &\simeq& 60~Z^{3/2}
       \left(\frac{\Mdot}{1.4\times 10^{-7} M_\sun \yr}\right)^{11/6} 
 \nonumber \\
   &\times&
		 \left(\frac{M_*}{M_\sun}\right)^{1/2}
	         \left(\frac{T}{10^3 K}\right)^{-3/2}
	         \left(\frac{\aMRI}{10^{-2}}\right)^{-4/3}
		 ~M_\earth.\label{eq:Mciso}
\end{eqnarray}
Note that this estimate should be considered as a maximum achievable size and that, more realistically, cores may only reach a fraction of $\Miso$. Therefore in the following section we select the core mass $M_c = f \Miso$.
The isolation mass of the core at $\acrit$
goes as $\Miso \propto \Mdot^{11/6} M_*^{1/2}$.  If we assume that $\Mdot$ is related to $M_*$ as discussed in $\S$\ref{sec:acrit} then $\Miso \propto
M_*^{2}$ to $M_*^{4}$.  
Therefore, super-earth cores form preferentially at the
inner boundary of the dead zone around more massive stars.

\subsection{Gas Giant Formation}
As our numerical algorithm does not include gas accretion onto the cores we can use the estimate of core mass derived in the previous section to look at the potential for further growth
using the approximations of \cite{IdaLin2005}. 
The growth of a planet of mass $M_p$ due to the gas capture can be approximated as
\begin{equation}
\frac{d M_p}{d t} \approx \frac{M_p}{\tau}
\end{equation}
where 
\begin{equation}
\tau \approx \tau_0\left(\frac{M_p}{M_\earth}\right)^{-3}
\end{equation}
with $\tau_0 \approx 10^{10}$ years (see \citet{IdaLin2005}, although $\tau_0$ is likely to be
smaller in the inner regions of the disk where $\rho_g$ is relatively
large).  As long at the final mass of the planet is much greater than
the original core mass, the timescale for giant planet formation is then
\begin{equation} 
t_{\rm giant} \approx
\frac{\tau_0}{3}\left(\frac{M_c}{M_\earth}\right)^{-3}.
\end{equation}

This inference effectively means that in order for a gas giant to form
within the lifetime of the evolving disk, the core mass must
at least be of the order of 10 $M_\earth$.  
Using the relationship derived 
in equation (\ref{eq:Mciso}) implies that $t_{\rm giant}\propto
\Mdot^{-11/2}M_*^{-3/2}$.  Folding in the relationship between $\Mdot$ and $M_*$ yields $t_{\rm giant}\propto M_*^{-7}$ to $M_*^{-12}$.  
This very steep function demonstrates that
giant planets will not have time to form at the inner edge of the dead
zone around less massive stars, but will form ubiquitously around higher mass stars
with higher accretion rates.

\subsection{Asymptotic Mass and Multiple Planet Systems}
The observationally inferred mass of the known planets around stars
with $M_\ast > 2 M_\odot$ is in the range of $2-20 M_J$ which is near
the upper end of the $M_p$ distribution of known planets around
solar-type stars.  In the core-accretion scenario, the asymptotic mass
of gas giants is determined by a thermal truncation condition
\citep{Lin.Papaloizou.1993} such that
\begin{equation}
M_p \simeq f_a (h/r)^3 M_\ast, 
\end{equation}
where the constant $f_a$ is of the order of 1 to 10 and depends on the detailed thermal structure of the disk \citep{DobbsDixon.etal.2007, IdaLin2005}. 
We use equation (\ref{eq:acrit}) to determine the value of $h/r =
c_s/v_K$ at $\acrit$.  The temperature and sound speed at the
inner edge of partially dead zone is, by construction, independent of $M_\ast$ and $\Mdot$.
We find 
from equations (\ref{eq:acrit}) and (\ref{eq:mdot})
that, at $\acrit$, $h/r \sim 0.05$ during the main phase of disk evolution.
From this result, we estimate that
\begin{equation}
M_p \approx 0.3 f_a \left(\frac{h/r}{0.05}\right)^3 \left(\frac{M_*}{2 M_\sun}\right) M_J.
\end{equation}

Figure 2 shows that the dependence of $\Mdot$ on $M_\ast$ in equation
(\ref{eq:mdot}) is an average relation with considerable dispersion.
Furthermore, $\Mdot$ declines with the stellar age $t_\ast$. Within our model
planets with larger $M_p$ on longer period orbits can form during the early
epochs of disk evolution when $\Mdot$ is much larger and $\acrit$ is correspondingly further out.
However, the rapid-accretion phase may be
too brief for embryos to reach their isolation mass. Nevertheless,
cores with sufficient mass to initiate efficient gas accretion are
likely to emerge when $\Mdot$ is reduced to that approximated by
equation (\ref{eq:mdot}) and when the timescale for core formation at
$\acrit$ becomes comparable to the disk evolution timescale.
Additionally, due to this spread in $\Mdot$ around stars with similar $M_\ast$
we anticipate there will be a wide distribution of
planetary masses, albeit the mean value of $M_p$ should vary with $M_\ast$.

In disks with protracted high-$\Mdot$ evolution, the first generation
of gas giants forms rapidly. Once the first planet has formed and
grown large enough to open a gap, the outer edge of this gap
provides another pressure maximum capable of trapping solids and promoting
the formation of the next planet (\cite{Bryden.etal.2000}).  
We anticipate a prolific production of multiple planetary systems for
intermediate-mass stars. 

\section{Summary and Discussion}\label{sec:summary}
While planets are likely to form by the same basic mechanisms
regardless of the environment
in which they form, the properties of their host stars and the detailed structure of their nascent disks will strongly
affect the statistical outcome of the formation process.  Observations hint that
planets may form systematically more efficiently around intermediate
mass stars, and that there is a statistically significant lack of giant planets on orbits
with semi-major axes much smaller than 1 AU.  In this paper we propose a mechanism which
forms giant planets preferentially around intermediate mass stars with radial
distributions roughly consistent with these observations.  
In this model the gaseous protoplanetary disk evolves due to MRI-driven turbulence, creating a pressure maximum at the inner edge of the dead zone ($\acrit$) which traps solid material.
In order for the cores formed at this location to grow large enough
to seed giant planets, the inner edge of the dead zone must be
sufficiently far from the host star. 
We demonstrate that, as $\acrit$ is roughly proportional to $M_*$,
this condition is only likely
to be met around intermediate mass stars.

The amount of solids which accumulates near $\acrit$ is limited by momentum feedback on the gas by the solids.
This is interesting as it means that 
the surface density of solids at $\acrit$ depends on the gas surface density and \emph{not} on the initial fraction of solids, except in extremely metal poor disks.  
This suggests that while there may be a critical metallicity required in order to form planets by the mechanism described in this paper, beyond that critical value the frequency of giant planets should depend only weakly on stellar metallicity.

The prolific production of gas giants near $\acrit$
can also promote the emergence of additional gas giants at
larger distances from the same host stars.  We may expect the fraction of
intermediate-mass stars with multiple Jupiter-mass planets is likely
to be larger than that around solar-type stars.  Nevertheless, we
anticipate the peak in the planets' semi-major axis distribution to be around 1
AU.  This corresponds to the location of the original
pressure maximum at the inner edge of the dead zone.  Quantitative
verification of this expectation requires population synthesis which
will be carried out and presented elsewhere.  Observational
confirmation of this peaked period distribution will provide clues and
constraints on the outstanding issue of magnetic turbulent transport
in protostellar disks.

\vspace{1em}
\noindent ACKNOWLEDGMENTS.  We thank D. Fischer, S. Ida, and P. Bodenheimer for many useful conversations.  This work is supported by NASA (NAGS5-11779,
NNG06-GF45G, NNX07A-L13G, NNX07AI88G), JPL (1270927), and
NSF(AST-0507424).

\appendix
\section{Functions I and J}
The functions $I$ and $J$ used in equation (\ref{eq:4}) and (\ref{eq:gas_vel}) are
\begin{eqnarray}
I(x) &=& \frac{\sqrt{2}}{4x}[f_1(x)+f_2(x)],\nonumber\\
J(x) &=& \frac{\sqrt{2}}{4x}[-f_1(x)+f_2(x)],\nonumber\\
f_1(x) &=& \frac{1}{2}\ln\left(\frac{x^2+x\sqrt{2}+1}{x^2-x\sqrt{2}+1}\right),\nonumber\\
f_2(x) &=& \arctan\left(x\sqrt{2}+1\right)+\arctan\left(x\sqrt{2}-1\right).
\end{eqnarray}
In the limit of $x \ll 1$, $I=1$ and $J=x^2/3$, and in the limit of $x \gg 1$, $I=J=\sqrt{2}\pi/(4x)\approx 1/x$.

%%%%%%%%%%%%%%%%%%%%%%%%%%%%%%%%%%%%%%%%%%%%%%%%%%%%%%%%%%%%%%%%%%
%\bibliography{copy}

%%%%%%%%%%%%%%%%%%%%%%%%%%%%%%%%%%%%%%%%%%%%%%%%%%%%%%%%%%%%%%%%%%
\clearpage

\begin{figure}
\begin{center}
% ~/Research/Scripts/IMS/simple_vertical/comp_structure.pro
\plotone{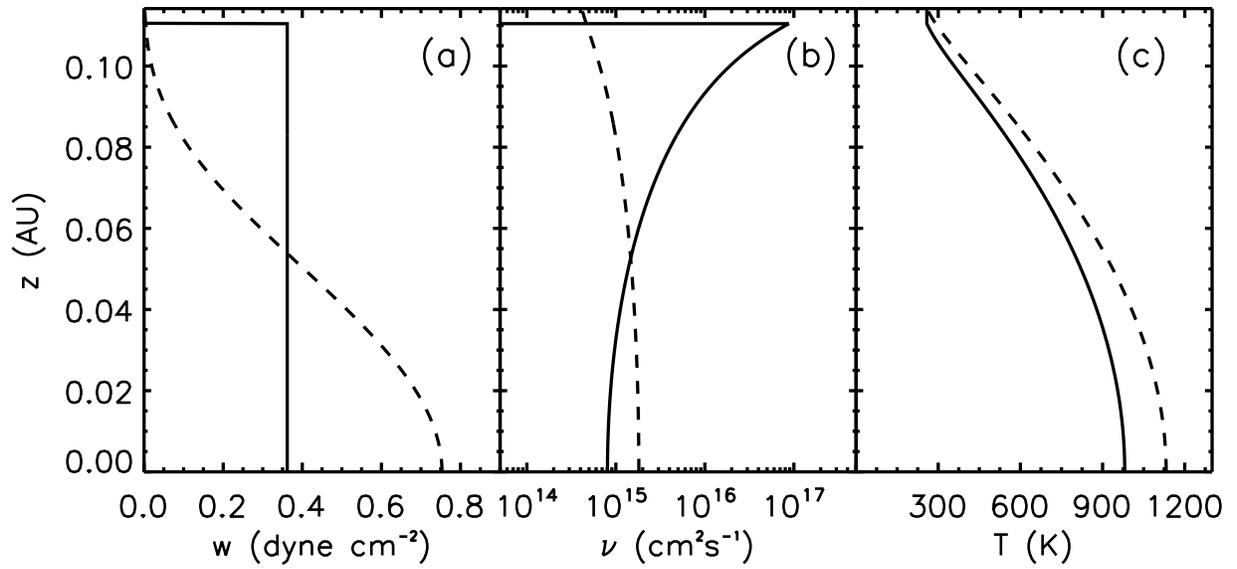}
\caption{A comparison of the disk structure resulting from the viscosity prescription from eq. (\ref{eq:nu})
(solid curves) to disks with the traditional ``$\alpha$-P formalism''
(dashed curves).  The panel (a) shows the
accretion stress, panel (b) the viscosity, and panel (c) the
temperature.}
   \label{zTemp}
\end{center}
\end{figure}
\clearpage

\begin{figure}
\begin{center}
% ~/Research/Scripts/IMS/acrit.pro
\plotone{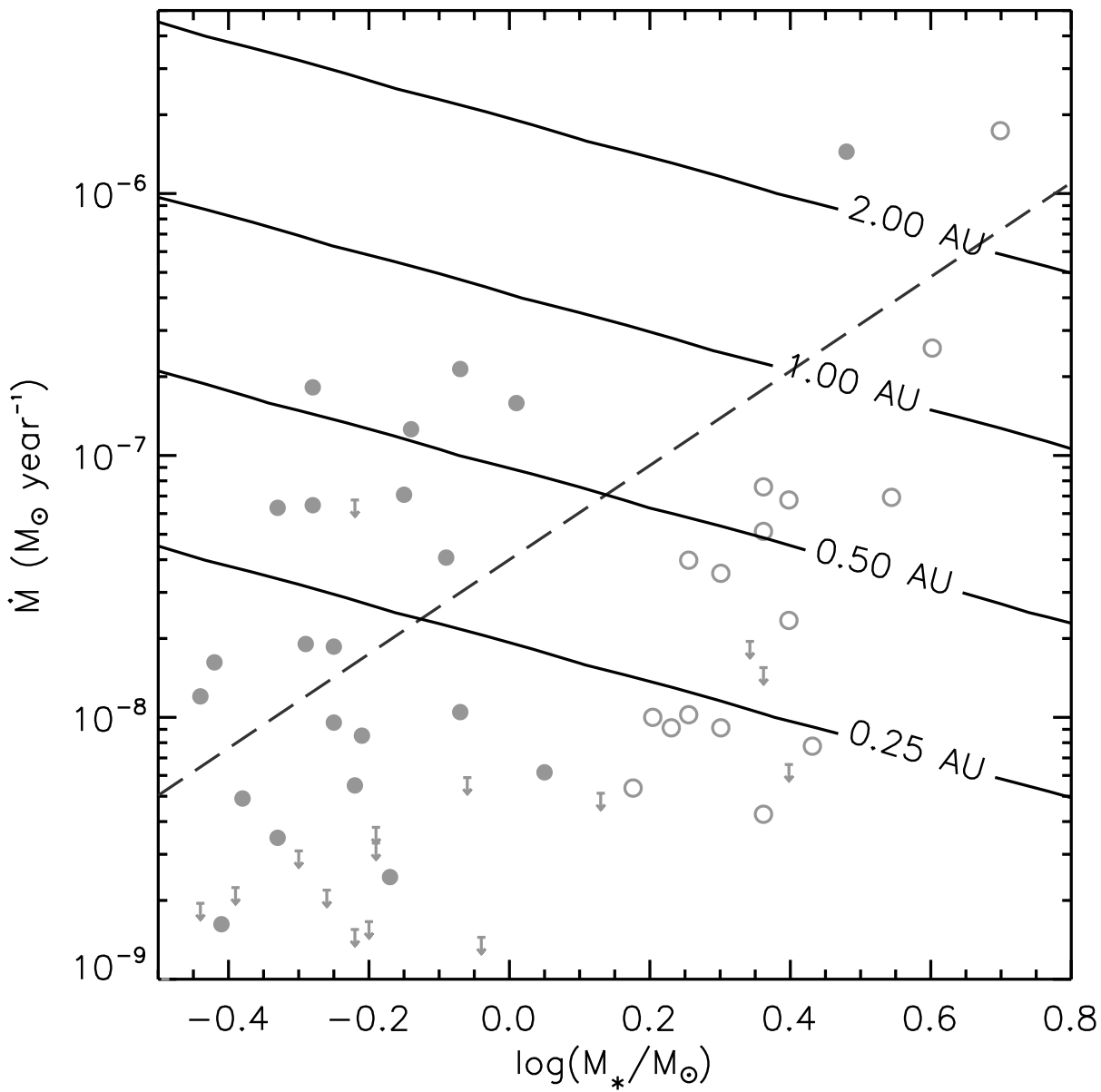}
\caption{The curves show the position of the inner edge of the
dead zone as a function of stellar mass and mass accretion rate.
Symbols represent observations of mass accretion rates for stars of
various masses (see text for details).  The dashed line shows the
best-fit for the $\rho$-Oph cluster from \cite{Natta.etal.2006}.}
   \label{acrit}
\end{center}
\end{figure}
\clearpage

\begin{figure}
\begin{center}
\plotone{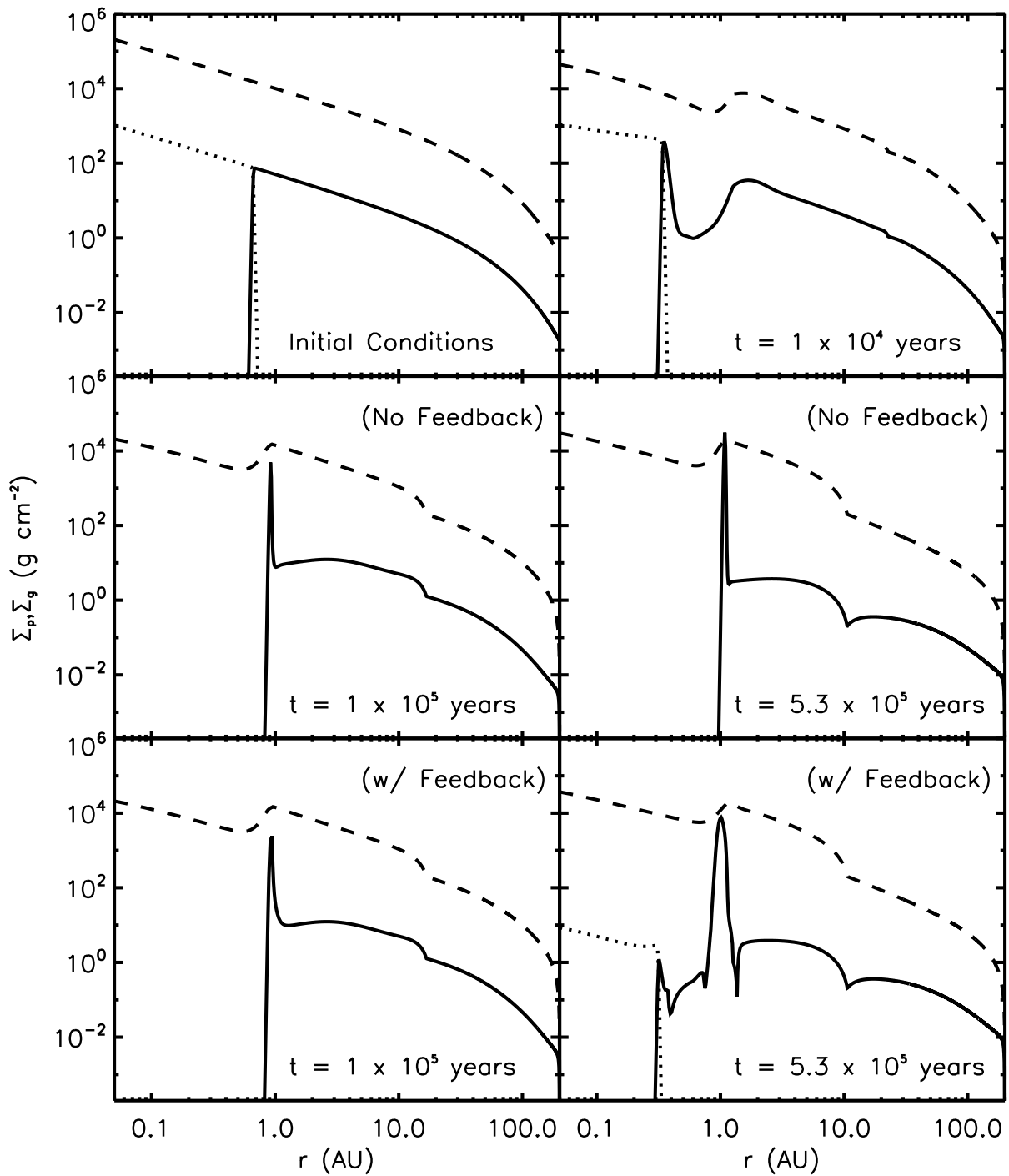}
\caption{Evolution of the surface density of solids (solid curve), gas (dashed curve) and heavy element in vapor form (dotted curve). The middle panels show the evolution neglecting feedback of the dust on the gas while the bottom panels include feedback.}\label{fig:sigma}
\end{center}
\end{figure}
\clearpage

\begin{figure}
\begin{center}
\plotone{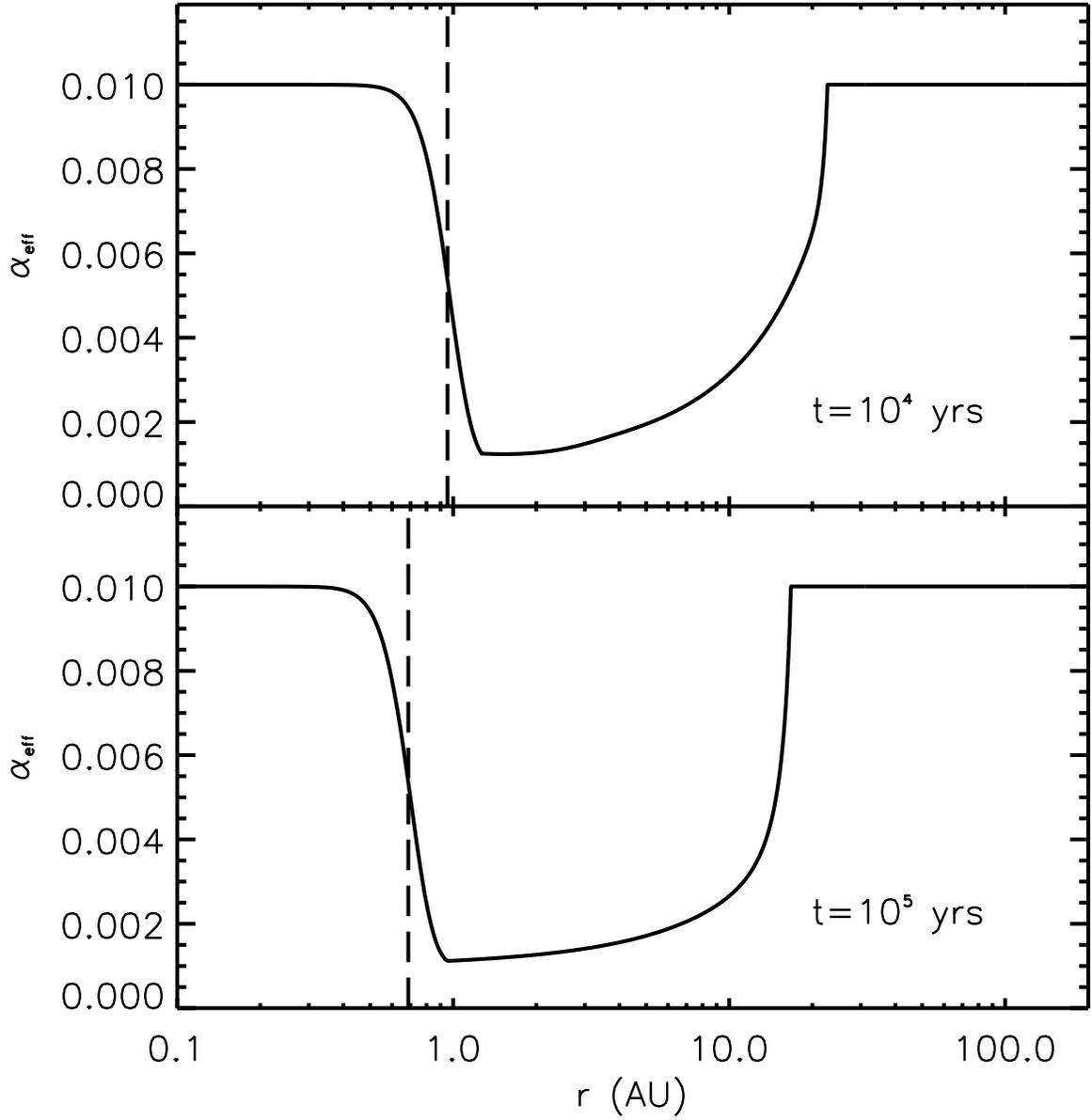}
\caption{The calculated $\aeff$ as a function of radius at $10^4$ and $10^5$~years.  The inner edge of the dead zone, $\acrit$ is indicated by the dotted line.}\label{fig:alpha}
\end{center}
\end{figure}
\clearpage

\begin{figure}
\begin{center}
\plotone{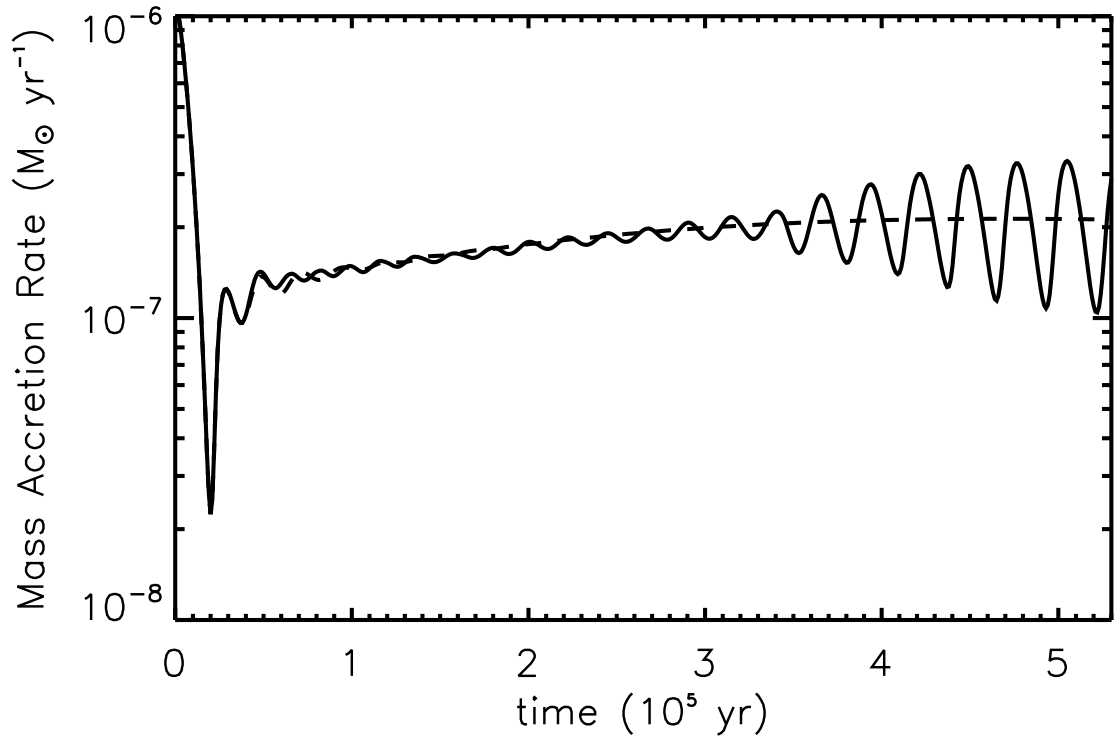}
\caption{The mass accretion rate onto the star as a function of time.  The solid curve indicates the mass accretion rate when feedback is included while the dashed curve neglects feedback.}\label{fig:Mdot}
\end{center}
\end{figure}
\clearpage

\end{document}